\documentclass[
aps,%
12pt,%
final,%
oneside,%
onecolumn,%
nobibnotes,%
nofootinbib,%
superscriptaddress,%
centertags,
floatfix,
secnumarabic,
notitlepage]%
{revtex4-1}%
\usepackage{graphicx}
\usepackage{amssymb,amsmath}
\usepackage{bm}
\usepackage{slashed}
\usepackage{subfigure}
\usepackage{multirow}
\RequirePackage[english]{babel}
\usepackage{hyperref}
\usepackage[normalem]{ulem}
\usepackage{xcolor}
\usepackage[draft]{changes}

\begin{document}

\selectlanguage{english}
\normalsize

\title{Associated charmonium-bottomonium production in a single boson $e^+e^-$~annihilation}

\author{\firstname{I.~N.}~\surname{Belov}}
\email{ilia.belov@cern.ch}
\affiliation{SINP MSU, Moscow, Russia}
\affiliation{Physics department of MSU, Moscow, Russia}

\author{\firstname{A.~V.}~\surname{Berezhnoy}}
\email{Alexander.Berezhnoy@cern.ch}
\affiliation{SINP MSU, Moscow, Russia}

\author{\firstname{E.~A.}~\surname{Leshchenko}}
\email{leshchenko.ea17@physics.msu.ru}
\affiliation{Physics department of MSU, Moscow, Russia}


\begin{abstract}
\small
The production cross sections of $J/\psi~\eta_b$, $\Upsilon\;\eta_c$ pairs in a single boson $e^+e^-$~annihilation have been studied in a wide range of energies, which will be achieved  at future $e^+e^-$~colliders. The main color singlet contributions to the production processes are taken into account, including the one loop QCD contribution.
\end{abstract}

\maketitle
\normalsize

\section{Introduction}

Despite its long history, the heavy quark physics continues to attract the attention of both theorists and experimentalists. Recently, due to the LHC experiments, the BELLE-II experiment and the BES-III experiment almost every year is marked by a discovery in this area. One of the hot topics of research is the quarkonium pair production. It is worth to remind here the intrigue related to the observation of $J/\psi~\eta_c$ pair in the $e^+e^-$ annihilation: the theoretical predictions ~\cite{Braaten:2002fi} underestimated  the real yield measured at BELLE and BaBar~\cite{Abe:2004ww,Aubert:2005tj} by the order of magnitude. This stimulated numerous studies~\cite{ee2jpsietac,ee2jpsietac_rwf},
as a result of which the fair agreement with the data has been achieved.
The previous year gave researchers a new sensation: the LHCb Collaboration observed the structure in the $J/\psi~J/\psi$ spectrum at large statistics~\cite{Aaij:2020fnh}.  This circumstance led to a real explosion of interest to this topic.
 
Looking to the future,  we  study the processes of the paired quarkonium production  which can not be observed at the existing experiments due to low interaction energy, namely, the  production of $J/\psi~\eta_b$ pairs  and  $\Upsilon~\eta_c$ pairs in  the  $e^+e^-$ annihilation. These processes can be investigated in the framework of several discussed projects:  ILC,  FCC, and the muon collider.  At  ILC and FCC, studies are planned at the energies of the order of $Z$-boson’s mass: the energy range announced for FCC is  $\sqrt{s} = 90\div400~\text{GeV}$~\cite{Koratzinos:2014cla} and $\sqrt{s} = 250~\text{GeV}$ is proposed for ILC~\cite{KEKInternationalWorkingGroup:2019spu}. In the project of muon collider,  it is planned to implement the  $\mu^+\mu^-$ collisions  at energies from 3 TeV to 14 TeV~\cite{Long:2020wfp}, which are beyond the energy range studied in this work.

It is worth to note that the decays of $Z$-boson  to  the charmonium and the bottomonium  may be of some interest for the experiments at the LHC, see~\cite{Sirunyan:2019lhe}.

In our previous studies we already have investigated the paired $B_c$ production~\cite{Berezhnoy:2016etd}, as well as the $J/\psi~J/\psi$ and  the $J/\psi~\eta_c$ pair production  around the $Z$  mass within the NLO approximation~\cite{Berezhnoy:2021tqb}.  We have found that for these processes the loop corrections essentially contribute to the cross section values. This result is in agreement with the studies of other research groups investigated the paired quarkonium production in the  $e^+e^-$ annihilation. Both  cases of  $J/\psi~\eta_b$ and $\Upsilon~\eta_c$ production are special, because the tree level diagrams with the single gluon exchange can not contribute to the production of $c\bar c$  and $b \bar b$ pairs in the color singlet states, and therefore the lowest order QCD contribution to these processes contains loops. This one-loop contribution is of the order of ${\cal O}(\alpha^2 \alpha_s^4)$, that is why it makes sense to investigate it with the purely electromagnetic $J/\psi~\eta_b$ and $\Upsilon~\eta_c$ production which is of the order of ${\cal O}(\alpha^4)$, since $ \alpha_s^2 \sim \alpha$. 
 
When studying these processes, one cannot ignore the discussion of the role of color octet contributions. Indeed, from one side the paired color octet production is  suppressed as $v_{c \bar c}^2v_{b \bar b}^2\sim 0.03$, where $v_{c \bar c}$ and $v_{b \bar b}$ are velocities of quarks inside charmonium and botommonium, correspondingly. From the other side, it has no additional suppression by $\alpha_s^2$,  going through the tree level QCD diagrams. Therefore a color octet mechanism could essentially contribute to the discussed quarkonium-pair production. It deserves a separate detailed consideration, which goes beyond the scope of the current study.

Another problem that is outside the scope of this study is the relativistic effects caused by the use of relativistic or relativized wave functions. It is known that accounting for such effects can crucially change the predictions, especially for the charmonium-pair production in the  $e^+e^-$ annihilation, where accounting for the relativistic effects   decreases the virtuality of the intermediate gluon, and therefore essentially increases the cross section value (see for example~\cite{ee2jpsietac_rwf}).  Another examples of influence of the relativistic effects on  the paired quarkonium production could be found in~\cite{BcBc_rwf}.  
For the processes discussed in this study we do not expect such a huge change of the cross section due to relativistic effects, as for the paired charmonium production, but,  of course, it does not eliminate the problem,  and such effects should be  thoroughly investigated.

It is worth to mention that the one loop QCD diagrams do not contribute to the production  of  $J/\psi~\Upsilon$-pair in a single boson $e^+e^-$-annihilation. In leading order such a process goes via the electroweak $Z$ boson decay only.
 As we planned the current work as a continuation of our previous studies on the one loop QCD corrections, in a sense, this process is slightly out of our interest.   However, we keep it in our consideration for comparison with $J/\psi~\eta_b$ and $\Upsilon~\eta_c$ production. Thus, the following processes are considered in this study:
\begin{equation}
    \begin{cases}
        e^{+}e^{-}\xrightarrow{\gamma^*,\ Z^*}~J/\psi~\eta_b,\\
        e^{+}e^{-}\xrightarrow{\gamma^*,\ Z^*}~\Upsilon~\eta_c,\\
        e^{+}e^{-}\xrightarrow{\quad\ Z^*}~J/\psi~\Upsilon.\\
    \end{cases}
 \end{equation}

\section{Calculation technique}\label{CALC}

The production of the pair of  charmonium and bottomonium in a single boson annihilation is constrained by several selection rules, which are discussed below.

The production of the $J/\psi~\Upsilon$ pairs is not allowed in the photon exchange, as well as it does not go  via the vector part of  $Z$ vertex due to the charge parity conservation. The $J/\psi~\Upsilon$ pairs are produced via the interaction with the axial part of  $Z$ vertex only. 
As concerned  the $\eta_b~\eta_c$ pair production, it  goes neither via the photon exchange, nor via the $Z$-boson exchange: the photon decay to the $\eta_b~\eta_c$ pair is forbidden due to the charge parity conservation, and the $Z$ decay to the $\eta_b~\eta_c$ pair is forbidden due to the combined $CP$ parity conservation. 
At the same time the vector-pseudoscalar (VP) pairs: $J/\psi~\eta_b$ and $\Upsilon~\eta_c$ are produced via the exchange of both the photon and the $Z$-boson. 
These selection rules were reproduced directly in our calculations.


As we already  pointed out in the Introduction the tree level diagrams describing the single gluon exchange do not contribute to the discussed processes,  and the lowest order QCD contribution to such processes contains loops.
This one-loop QCD contribution is of the order of ${\cal O}(\alpha^2 \alpha_s^4)$ and therefore it could be  comparable with the pure electroweak tree level contribution, which is of the order of ${\cal O}(\alpha^4) \sim {\cal O}(\alpha^2 \alpha_s^4)$. 
Thus, when studying these processes, one should take into account  the 
 electroweak  contribution (EW) of the order of ${\cal O}(\alpha^4)$, the  one loop QCD contribution of the order of $ {\cal O}(\alpha^2 \alpha_s^4)$, and the  interference between them of the order of $\mathcal{O}(\alpha^{3}\alpha_{s}^{2})$: 
\begin{equation}\label{general}
|\mathcal{A}|^{2}=|\mathcal{A}^{EW}|^2 + 2Re(\mathcal{A}^{EW}\mathcal{A}^{QCD *}) + |\mathcal{A}^{QCD}|^2. 
\end{equation}

For the more detailed study of the processes we consider the amplitudes with different intermediate bosons separately:
\begin{multline}\label{eq:full_square}
    |\mathcal{A}|^{2}=|\mathcal{A}^{EW}_\gamma|^{2}+|\mathcal{A}^{EW}_Z|^{2}+2Re(\mathcal{A}^{EW}_\gamma \mathcal{A}^{EW *}_Z)+\\
    +2Re(\mathcal{A}^{EW}_\gamma \mathcal{A}^{QCD *}_\gamma)+2Re(\mathcal{A}^{EW}_Z \mathcal{A}^{QCD *}_Z)+2Re(\mathcal{A}^{EW}_\gamma \mathcal{A}^{QCD *}_Z)+2Re(\mathcal{A}^{EW}_Z \mathcal{A}^{QCD *}_\gamma)+\\
    +|\mathcal{A}^{QCD}_\gamma|^{2}+|\mathcal{A}^{QCD}_Z|^{2}+2Re(\mathcal{A}^{QCD}_\gamma \mathcal{A}^{QCD *}_Z).
\end{multline}


The production of double heavy quarkonia  is  described in the framework of nonrelativistic QCD (NRQCD). This formalism allows to factor out the perturbative degrees of freedom and therefore separate the production mechanism into  hard and soft subprocesses, using the hierarchy of scales for the quarkonia, which is $m_q >> m_q v, \ m_q v^2, \Lambda_{QCD}$, where $m_q$ is  the heavy quark mass and $v$ is  the velocity of  heavy quark in the quarkonium. The hard subprocess  corresponds to the perturbative production of $q\bar q$-pair, while the soft subprocess  corresponds to the fusion of quarks into the bound state.


To compute the matrix elements for the studied  processes, we start from the matrix element for $e^+e^-\to c(p_c) \bar c(p_{\bar c}) b(p_b) \bar b(p_{\bar b})$ with heavy quarks and antiquarks on their mass shells: $p_c^2=p_{\bar c}^2=m_c^2$ and $p_b^2=p_{\bar b}^2=m_b^2$. As we put $v=0$ before the projection  onto the bound states, the momentum $P$ of  charmonium and the momentum
$Q$ of bottomonium are related with the heavy quark momenta as follows:
\begin{align}\label{momenta}
&J/\psi,\ \eta_c\ \begin{cases} 
&p_c = P/2 \\ &p_{\bar c} = P/2 \\
\end{cases}
&\Upsilon,\ \eta_b\ \begin{cases}
&p_b = Q/2 \\ &p_{\bar b} = Q/2
\end{cases}
\end{align}

To construct the bound states we replace the spinor products $v(p_{\bar q})\bar u(p_q)$ by the appropriate covariant projectors for color-singlet spin-singlet and spin-triplet
states:

\begin{align}
\label{eq:projectors_cc}
     &\Pi_{J/\psi}(P,m_{c})=\frac{\slashed P-2 m_{c}}{2\sqrt{2 m_{c}}}\ \slashed \epsilon^{J/\psi} \otimes \frac{\boldsymbol 1}{\sqrt{N_c}},     &\Pi_{\eta_{c}}(P,m_{c})=\frac{\slashed P-2 m_{c}}{2\sqrt{2 m_{c}}}\gamma^{5}\otimes \frac{\boldsymbol 1}{\sqrt{N_c}},\\
\label{eq:projectors_bb}
     &\Pi_{\Upsilon}(Q,m_{b})=\frac{\slashed Q-2 m_{b}}{2\sqrt{2 m_{b}}} \ \slashed \epsilon^\Upsilon \otimes \frac{\boldsymbol 1}{\sqrt{N_c}},
     &\Pi_{\eta_{b}}(Q,m_{b})=\frac{\slashed Q-2 m_{b}}{2\sqrt{2 m_{b}}}\gamma^{5}\otimes \frac{\boldsymbol 1}{\sqrt{N_c}}, 
 \end{align}
where $\epsilon^{J/\psi}$ and  $\epsilon^{\Upsilon}$ are the polarizations of the  $J/\psi$ and  $\Upsilon$ mesons, satisfying the following constraints: $\epsilon^{J/\psi}\cdot {\epsilon^{J/\psi}}^*=-1$, $\epsilon^{J/\psi}\cdot P=0$, $\epsilon^{\Upsilon}\cdot {\epsilon^{\Upsilon}}^*=-1$ and $\epsilon^{\Upsilon}\cdot Q=0$.

\begin{figure}[ht]  
 \centering 
 \subfigure[]{
\includegraphics[width=0.32\linewidth]{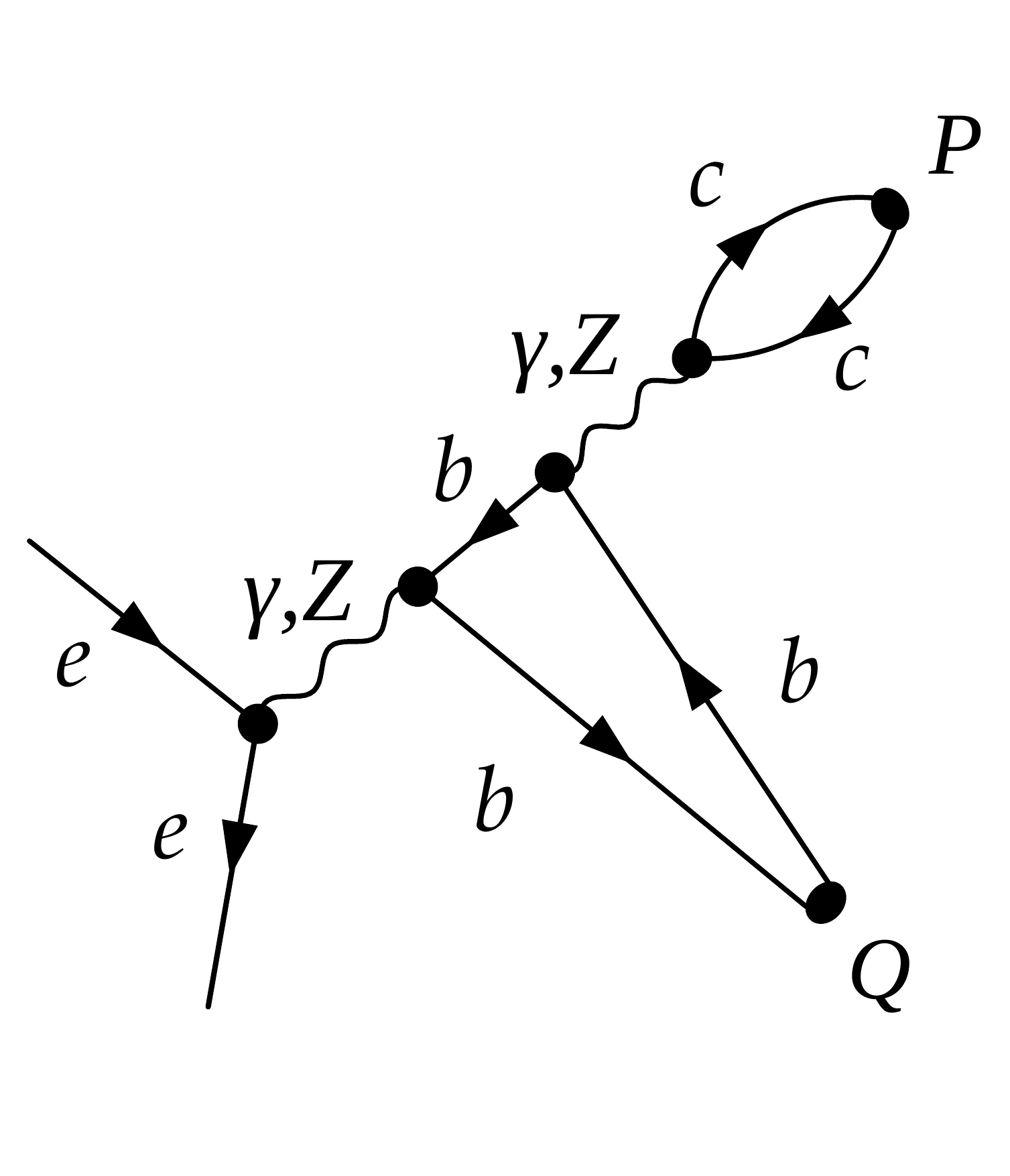} \label{fig:picQED}}
\hfill
\subfigure[]{
\includegraphics[width=0.32\linewidth]{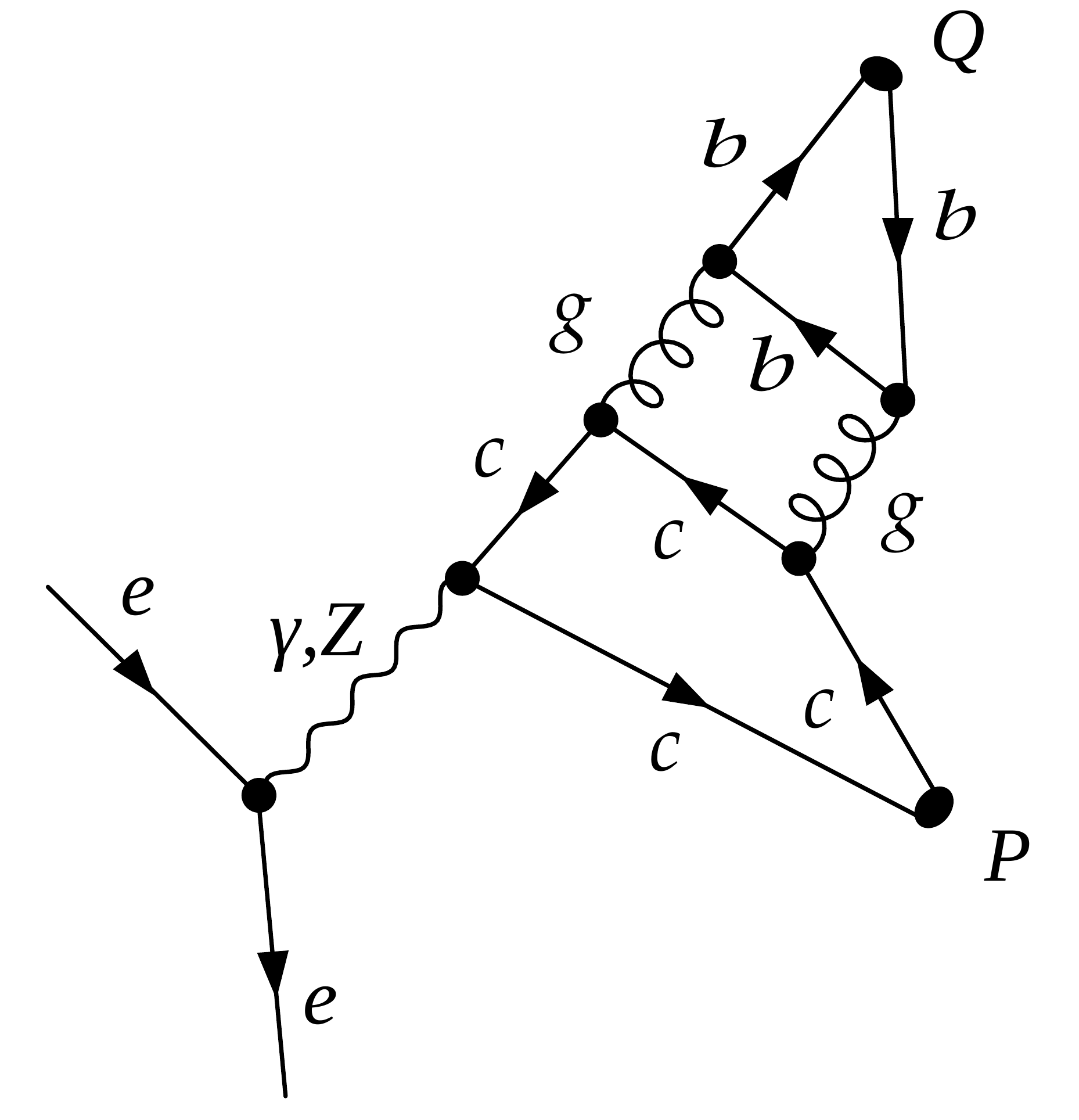} \label{fig:picQCD1}} 
\hfill
\subfigure[]{
\includegraphics[width=0.28\linewidth]{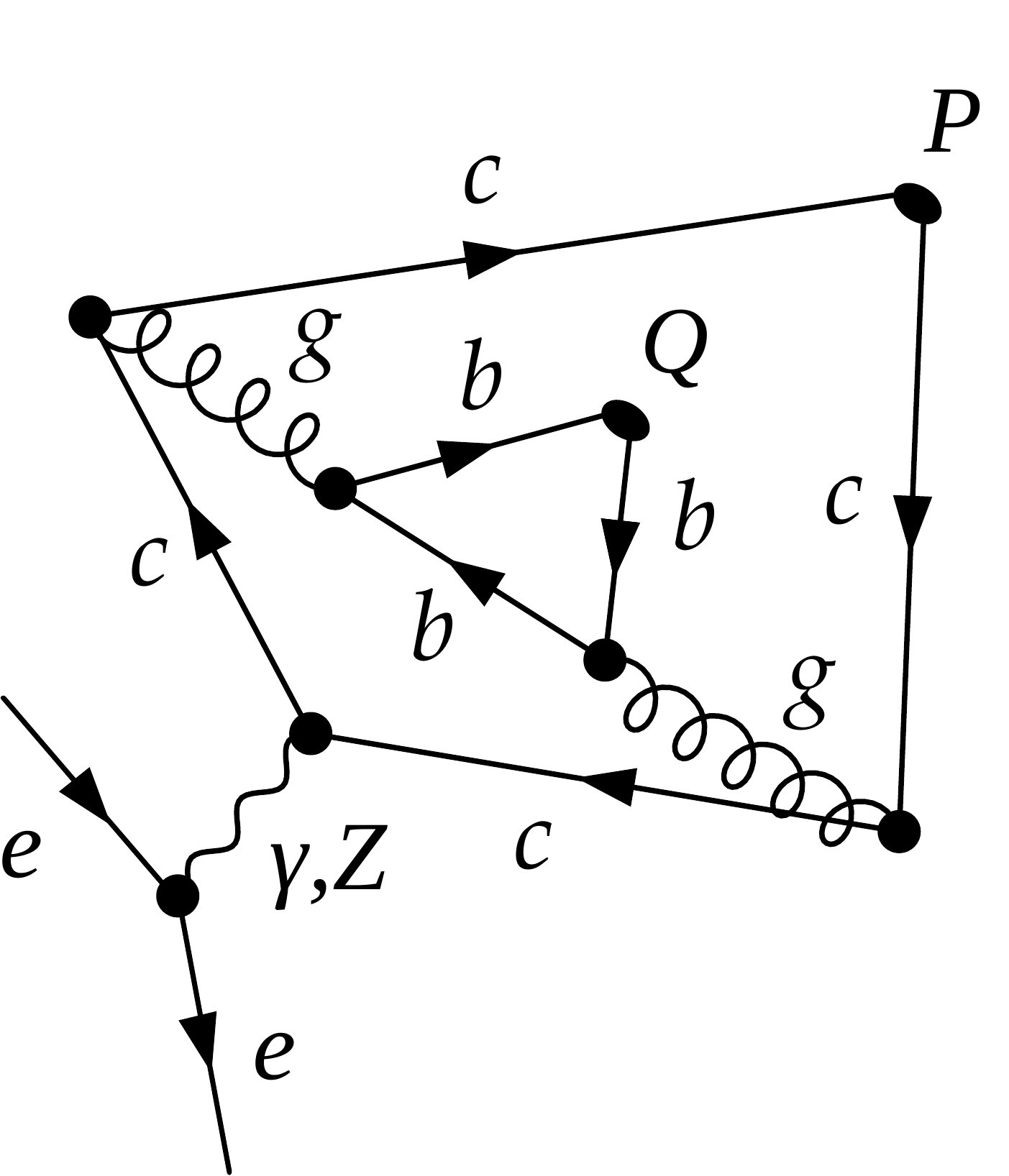} \label{fig:picQCD2}}
\caption{The diagram examples contributing to the  $e^+e^-\to\ J/\psi\ \eta_b$ process: the tree level electroweak diagram (a) ;  the diagrams with a QCD loop ( (b)~and~(c) ).}
\label{fig:diagrams}
\end{figure}

These operators close the fermion lines into traces.
The examples of diagrams contributing to the process $e^+e^-\rightarrow J/\psi~\eta_b$ are shown in Figure~\ref{fig:diagrams}.

The factorized matrix elements have the following form:

\begin{equation}
    \mathcal{A}\left(e^+e^-  \to J/\psi \: \eta_{b}\right) = \frac{\langle O_{J/\psi} \rangle^{1/2}\langle O_{\eta_b} \rangle^{1/2} }{N_c}\mathcal{M}_{J/\psi\:\eta_{b}}^\mu \epsilon^{J/\psi}_\mu,
\end{equation}
\begin{equation}
    \mathcal{A}\left(e^+e^- \to \Upsilon \: \eta_{c}\right) =\frac{\langle O_{\Upsilon} \rangle^{1/2}\langle O_{\eta_c} \rangle^{1/2} }{N_c} \mathcal{M}_{\Upsilon\:\eta_{c}}^\mu \epsilon^\Upsilon_\mu,
\end{equation}
\begin{equation}
    \mathcal{A}\left(e^+e^- \to J/\psi \: \Upsilon\right) =\frac{\langle O_{J/\psi} \rangle^{1/2}\langle O_{\Upsilon} \rangle^{1/2} }{N_c} \mathcal{M}_{J/\psi\:\Upsilon}^{\mu\nu}\epsilon^{J/\psi}_\mu \epsilon^\Upsilon_\nu,
\end{equation}
where $\mathcal{M}_{J/\psi\:\eta_{b}}^\mu$, $\mathcal{M}_{\Upsilon\:\eta_{c}}^\mu$, and  $\mathcal{M}_{J/\psi\:\Upsilon}^{\mu\nu}$ are the hard production amplitudes of two quark-antiquark pairs projected onto the quark-antiquark states with zero relative velocities  and the appropriate quantum numbers
by projectors \eqref{eq:projectors_cc} and \eqref{eq:projectors_bb}. The NRQCD matrix elements $\langle O_{\Upsilon} \rangle$,  $\langle O_{J/\psi} \rangle$, $\langle O_{\eta_b} \rangle$ and $\langle O_{\eta_c} \rangle$ are vacuum-saturated analogs of the NRQCD matrix elements $\langle O (^3S_1)\rangle$  and $\langle O (^1S_0)\rangle$ for annihilation decays defined in \cite{Bodwin:1994jh}.
The numerical values of these matrix elements can be estimated  from the experimental data on decays~\cite{Braaten:2002fi,Bodwin:2007fz,Chung:2010vz}, or adopted from the potential models, such as~\cite{Eichten:2019hbb}, using 
the relation $\langle O \rangle \approx \frac{N_c}{2\pi} |R(0)|^2$, where $R(0)$ is the quarkonium wave function at origin (see Table~\ref{tab:nrqcd_me} of  Appendix~\ref{APP1}).

\section{Workflow}
 The diagrams and the corresponding analytic expressions are generated with the   \texttt{FeynArts}-package~\cite{Hahn:2000kx} in Wolfram Mathematica.  The electroweak contribution to the production amplitudes is determined by the tree diagrams of type~(a) shown in Figure~\ref{fig:diagrams}, whereas the one-loop diagrams of type~(b) and~(c)  contribute to the QCD amplitudes.

  We obtain  4 nonzero electroweak and 6 nonzero QCD diagrams for each subprocesses $e^+e^- \xrightarrow{\gamma^*} J/\psi\ \eta_b$,  $e^+e^- \xrightarrow{Z^*} J/\psi\ \eta_b$, $e^+e^-\xrightarrow{\gamma^*} \Upsilon\ \eta_c$ and $e^+e^- \xrightarrow{Z^*} \Upsilon\ \eta_c$. 
 The associative production of two vector states $J/\psi$ and $\Upsilon$ is described only by the tree electroweak diagrams of type~(a). 
 These results, as well as the explicitly obtained zero contribution to the process of $\eta_{b}\ \eta_{c}$ production, are in exact agreement with the earlier discussed selection rules,  providing the additional verification of the procedure. 
 
To calculate the tree level amplitudes we use only \texttt{FeynArts}~\cite{Hahn:2000kx} and \texttt{FeynCalc}~\cite{Shtabovenko:2020gxv} packages in \texttt{Wolfram Mathematica}, 
while the computation of the one loop amplitudes requires a more complicated toolchain: \texttt{FeynArts} $\rightarrow$ \texttt{FeynCalc}(\texttt{TIDL}) $\rightarrow$ texttt{Apart}~\cite{Feng:2012iq} $\rightarrow$ \texttt{FIRE}~\cite{Smirnov:2008iw}$ \rightarrow$ \texttt{X}~\cite{Patel:2016fam}.

All necessary algebraic calculations with Dirac and colour matrices, including the trace evaluation, are done within the \texttt{FeynCalc} package.  At the next step the Passarino-Veltman reduction is carried out using the  \texttt{TIDL} library implemented in \texttt{FeynCalc}. The \texttt{Apart} function does the extra simplification by partial fractioning for IR-divergent integrals. The \texttt{FIRE} package provides the complete reduction of  the  integrals obtained in the previous stages  to master integrals. This package  implements several strategies for the IBP reduction mostly based on the Laporta algorithm~\cite{Laporta:2001dd}.  The master integrals are then evaluated by substitution of their analytical expressions with the help of \texttt{X}-package.

The conventional dimensional regularization (CDR) scheme with  $D$-dimensional loop and external momenta is used to compute the  QCD amplitudes. 
Each QCD amplitude for the separate loop diagram carries a singular term of the order of ${\cal O}(1/\varepsilon)$. These terms  cancel out after summing over the set of QCD amplitudes,  as expected.

It is known that $\gamma^{5}$  is poorly defined  in $D$-dimensions. In the current study the so-called naive interpretation of $\gamma^5$ was used: $\gamma^5$ anticommutes with all other matrices and therefore disappears in traces with an even number of $\gamma^5$. In traces with an odd number of $\gamma^5$ the remaining $\gamma^5$ is moved to the right and replaced by
\begin{equation}\label{g5} 
    \gamma^{5}=\frac{-i}{24}\varepsilon_{\alpha\beta\sigma\rho}\gamma^{\alpha}\gamma^{\beta}\gamma^{\sigma}\gamma^{\rho}.
\end{equation}
Since $\varepsilon_{\alpha\beta\sigma\rho}$ is contracted after the regularization procedure,  we can safely treat it as $4$-dimensional.

\section{Analytical form of the amplitudes}

The relative simplicity of the considered processes makes it possible to provide the analytical expressions for the amplitudes right in the text.

The electroweak amplitudes for the processes 
$e^+e^- \rightarrow J/\psi~\eta_b$ and $e^+e^- \rightarrow \Upsilon~\eta_c$ can be written as per

\begin{multline}\label{eq:amp_ew_c}
\mathcal{A}_{EW}\left(e^+e^-\to J/\psi~\eta_b\right) = \frac{\langle O_{J/\psi} \rangle^{1/2}\langle O_{\eta_b} \rangle^{1/2} }{N_c} \times
\\
\times\frac{-3~e^4~e_b~e_c}{2 m_c\sqrt{m_b m_c}~\left(s+4m_c^2-4m_b^2\right)} \left(b_{\gamma}J_{\mu} + b_Z\widetilde{J}_{\mu}\right)\epsilon_{\nu}^{J/\psi}P_{\rho}Q_{\sigma}\varepsilon^{\mu\nu\rho\sigma}.
\end{multline}

\begin{multline}\label{eq:amp_ew_b}
\mathcal{A}_{EW}\left(e^+e^-\to \Upsilon~\eta_c\right) = \frac{\langle O_{\Upsilon} \rangle^{1/2}\langle O_{\eta_c} \rangle^{1/2} }{N_c} \times \\
\times\frac{3~e^4~e_b~e_c}{2 m_b\sqrt{m_b m_c}~\left(s+4m_b^2-4m_c^2\right)} \left(c_{\gamma}J_{\mu} + c_Z\widetilde{J}_{\mu}\right)\epsilon_{\nu}^{\Upsilon}P_{\rho}Q_{\sigma}\varepsilon^{\mu\nu\rho\sigma},
\end{multline}
where $e_c=2/3$ and $e_b=-1/3$ are the quark charges,  $J_{\mu}$ and $\widetilde{J}$
are the parts of electroweak current which describe the $e^+e^-$ annihilation to the virtual photon and to the virtual $Z$ boson correspondingly ($J_{\mu} = -i~\overline{e}\gamma_{\mu}e$, $\widetilde{J}_{\mu} = -i~\overline{e}^{~}\Gamma_{\mu}^{Z}e$), and 
\begin{align}\label{ckoeff}
&c_{\gamma} = \frac{4e_c}{s},
&c_Z = \left(\frac{4e_c\sin^2\theta_w-1}{\cos\theta_w\sin\theta_w}\right)\frac{1}{s-M_Z^2 + i\Gamma M_Z},\\ \label{bkoeff}
&b_{\gamma} = \frac{4e_b}{s},
&b_Z = \left(\frac{4 e_b\sin^2\theta_w+1}{\cos\theta_w\sin\theta_w}\right)\frac{1}{s-M_Z^2+i\Gamma M_Z}.
\end{align}

Such a simple structure of amplitudes (\ref{eq:amp_ew_c}) and  (\ref{eq:amp_ew_b}) is explained by the fact that  
only the vector, i.~e. the photon-like part of $Z q \bar q$ vertex contributes to the decays $Z^*\rightarrow J/\psi~\eta_b$ and  $Z^*\rightarrow \Upsilon~\eta_c$. 

The  analytical expressions for the cross sections of the electromagnetic production of $J/\psi~\eta_b$ and $\Upsilon~\eta_c$ pairs via the photon exchange are very simple and thus might be shown in the text:
\begin{multline}\label{sig_qed_jpsi_etab}
\sigma_{QED}\left(e^+e^-\xrightarrow{\gamma^*} J/\psi~\eta_b\right)  = \\ = \frac{32 \ \pi ^3 \alpha ^4 e_b^4 e_c^2 {\cal O}_{\eta_b} {\cal O}_{J/\psi} \left(s-4(m_b+m_c)^2\right)^{3/2} \left(s-4(m_b-m_c)^2\right)^{3/2}}{3 \ m_b m_c^3 s^3 \left(s-4 m_b^2+4 m_c^2\right)^2},
\end{multline}
\begin{multline}\label{sig_qed_upsilon_etac}
\sigma_{QED}\left(e^+e^-\xrightarrow{\gamma^*} \Upsilon~\eta_c\right)  = \\ = \frac{32 \ \pi ^3 \alpha ^4 e_c^4 e_b^2 {\cal O}_{\eta_c} {\cal O}_{\Upsilon} \left(s-4(m_b+m_c)^2\right)^{3/2} \left(s-4(m_b-m_c)^2\right)^{3/2}}{3 \ m_c m_b^3 s^3 \left(s-4m_c^2+4 m_b^2\right)^2}.
\end{multline}


It should be noted that in the $e^+e^- \rightarrow J/\psi~\eta_b$ and $e^+e^- \rightarrow \Upsilon~\eta_c$ processes the virtual photon transforming into the vector meson  ($\gamma^* \to J/\psi$ or  $\gamma^* \to \Upsilon$, see picture~(a) in Figure~\ref{fig:diagrams}) can be complemented by the analogous processes with  $Z$ boson. Formally, all theses processes  are of the same order in terms of  coupling constants.  However the latter ones are extremely suppressed by the $Z$ boson propagator as factors $\left(\frac{m_c^2}{M_Z^2-4m_c^2}\right)$ and $\left(\frac{m_b^2}{M_Z^2-4m_b^2}\right)$, correspondingly, and can be neglected. 
The analytical expressions for these contributions  can be found in ~\eqref{appendix_ew_c} and~\eqref{appendix_ew_b} of Appendix~\ref{APP2}.

 The QCD one loop contributions to the amplitudes of the  discussed processes have exactly the same Lorentz structure as  the electroweak contributions (\ref{eq:amp_ew_c}) and  (\ref{eq:amp_ew_b}):
 \begin{multline}\label{eq:amp_qcd_c}
\mathcal{A}_{QCD}\left(e^+e^-\to J/\psi~\eta_b\right) =
\frac{\langle O_{J/\psi} \rangle^{1/2}\langle O_{\eta_b} \rangle^{1/2} }{N_c} \times\\
\times \frac{4C_A C_F}{N_c}\sqrt{\frac{m_c}{m_b}}~e^2g_s^4~\Bigl( c_{\gamma}J_{\mu}+c_Z\widetilde{J}_{\mu}\Bigr) \epsilon_{\nu}^{J/\psi}P_{\rho}Q_{\sigma}~\varepsilon^{\mu\nu\rho\sigma}
   \times \\ \times \left(\frac{4 i m_{b}^2 C_0\left(4 m_{b}^2,m_{c}^2,2   m_{b}^2-m_{c}^2+\frac{s}{2};0,0,m_{c}\right)}{16   m_{b}^4-\left(s-4 m_{c}^2\right)^2}\right. +\\-   \frac{4 im_{c}^2 \left(-4 m_{b}^2+4 m_{c}^2-s\right)
   C_0\left(m_{c}^2,2
   m_{b}^2-m_{c}^2+\frac{s}{2},s;m_{c},0,m_{c}\right)}{64 m_{b}^6-16 m_{b}^4 \left(4 m_{c}^2+3
   s\right)-4 m_{b}^2 \left(16 m_{c}^4+8 m_{c}^2 s-3
   s^2\right)+\left(4 m_{c}^2-s\right)^3}+\\-\frac{\pi  \left(4
   m_{b}^2-4 m_{c}^2+3 s\right)}{64 m_{b}^6-16
   m_{b}^4 \left(10 m_{c}^2+s\right)+4 m_{b}^2
   \left(32 m_{c}^4-12 m_{c}^2 s-s^2\right)-\left(2
   m_{c}^2-s\right) \left(s-4 m_{c}^2\right)^2}+\\+\frac{i
   \left(4 m_{b}^2-4 m_{c}^2+3 s\right) \ln
   \left(\frac{2 m_{c}^2}{4 m_{b}^2-4
   m_{c}^2+s}\right)}{64 m_{b}^6-16 m_{b}^4 \left(10
   m_{c}^2+s\right)+4 m_{b}^2 \left(32 m_{c}^4-12
   m_{c}^2 s-s^2\right)-\left(2 m_{c}^2-s\right)
   \left(s-4 m_{c}^2\right)^2}+\\+\left. \frac{4 i \sqrt{s \left(s-4
   m_{c}^2\right)} \ln \left(\frac{\sqrt{s \left(s-4
   m_{c}^2\right)}+2 m_{c}^2-s}{2
   m_{c}^2}\right)}{-64 m_{b}^6+16 m_{b}^4 \left(12
   m_{c}^2+s\right)+4 m_{b}^2 \left(-48 m_{c}^4+8
   m_{c}^2 s+s^2\right)+\left(4
   m_{c}^2-s\right)^3}\right),
\end{multline}

\begin{multline}\label{eq:amp_qcd_b}
\mathcal{A}_{QCD}\left(e^+e^-\to \Upsilon~\eta_c\right) =
\frac{\langle O_{\Upsilon} \rangle^{1/2}\langle O_{\eta_c} \rangle^{1/2} }{N_c} \times\\
\times \frac{4 C_A C_F}{N_c}\sqrt{\frac{m_b}{m_c}}~e^2g_s^4~\Bigl(b_{\gamma}J_{\mu} + b_Z\widetilde{J}_{\mu}\Bigr) \epsilon_{\nu}^{\Upsilon}P_{\rho}Q_{\sigma}~\varepsilon^{\mu\nu\rho\sigma}\times\\
   \times \Bigg(\frac{4 i
   m_{c}^2 C_0\left(m_{b}^2,4 m_{c}^2,-m_{b}^2+2
   m_{c}^2+\frac{s}{2};m_{b},0,0\right)}{16
   m_{b}^4-8 m_{b}^2 s-16 m_{c}^4+s^2}+\\+\frac{4 i
   m_{b}^2 \left(4 m_{b}^2-4 m_{c}^2-s\right)
   C_0\left(m_{b}^2,-m_{b}^2+2
   m_{c}^2+\frac{s}{2},s;m_{b},0,m_{b}\right)}{64
   m_{b}^6-16 m_{b}^4 \left(4 m_{c}^2+3 s\right)-4
   m_{b}^2 \left(16 m_{c}^4+8 m_{c}^2 s-3
   s^2\right)+\left(4 m_{c}^2-s\right)^3}+\\-\frac{\pi 
   \left(-4 m_{b}^2+4 m_{c}^2+3 s\right)}{32
   m_{b}^6-32 m_{b}^4 \left(4 m_{c}^2+s\right)+2
   m_{b}^2 \left(80 m_{c}^4+24 m_{c}^2 s+5
   s^2\right)-\left(s-4 m_{c}^2\right)^2 \left(4
   m_{c}^2+s\right)}+\\-\frac{i \left(4 m_{b}^2-4
   m_{c}^2-3 s\right) \ln \left(\frac{2 m_{b}^2}{4
   m_{c}^2-4 m_{b}^2+s}\right)}{32 m_{b}^6-32
   m_{b}^4 \left(4 m_{c}^2+s\right)+2 m_{b}^2
   \left(80 m_{c}^4+24 m_{c}^2 s+5 s^2\right)-\left(s-4
   m_{c}^2\right)^2 \left(4 m_{c}^2+s\right)}+\\-\left. \frac{4 i
   \sqrt{s \left(s-4 m_{b}^2\right)} \ln
   \left(\frac{\sqrt{s \left(s-4 m_{b}^2\right)}+2
   m_{b}^2-s}{2 m_{b}^2}\right)}{64 m_{b}^6-48
   m_{b}^4 \left(4 m_{c}^2+s\right)+4 m_{b}^2
   \left(48 m_{c}^4+8 m_{c}^2 s+3 s^2\right)-\left(s-4
   m_{c}^2\right)^2 \left(4
   m_{c}^2+s\right)}\right),
\end{multline}
where $C_0$ is the scalar three-point Passarino-Veltman function $\mathtt{ScalarC0[s_1,s_{12},s_2;m_0,m_1,m_2]}$ defined in Package-\texttt{X}. 

As already mentioned the virtual photon does not decay  to the $J/\psi\: \Upsilon$ pair due to the charge parity conservation, while the virtual $Z$-boson does.  The Lorentz structure of this amplitude is a little bit more complicated than presented in (\ref{eq:amp_ew_c}) and (\ref{eq:amp_ew_b}), as  contains the additional polarization vector and consists of two components:
\begin{multline}\label{eq:amp_ew_bc}
{\cal A}_{EW}\left(e^+e^- \rightarrow J/\psi~\Upsilon\right) = \frac{\langle O_{J/\psi} \rangle^{1/2}\langle O_{\Upsilon} \rangle^{1/2} }{N_c} \times\\
\times \frac{3~e^4~e_be_c \sqrt{m_b m_c}}{\cos\theta_w\sin\theta_w \left(s-M_Z^2+i\Gamma M_Z\right)}
~\widetilde{J}_{\mu}\epsilon_{\nu_1}^{J/\psi}\epsilon_{\nu_2}^{\Upsilon}\varepsilon^{\mu \nu_1\nu_2\sigma}\times
\\
\times
\left(\frac{P_{\sigma}}{m_c^2 \left(4m_b^2-4 m_c^2-s\right)} - \frac{Q_{\sigma}}{m_b^2 \left(4 m_b^2-4
   m_c^2+s\right)}\right). 
\end{multline}

The photonic parts of the electroweak amplitudes~\eqref{eq:amp_ew_c} and \eqref{eq:amp_ew_b}
turn into each other under simultaneous permutations $m_b \longleftrightarrow m_c$, $e_b \longleftrightarrow e_c$ and $P \longleftrightarrow Q$, as expected (as well as cross sections~\eqref{sig_qed_jpsi_etab} and \eqref{sig_qed_upsilon_etac}). The same applies for the photonic parts of the one loop QCD amplitudes \eqref{eq:amp_qcd_c} and \eqref{eq:amp_qcd_b}.

It is interesting to note, that  EW and QCD amplitudes for the VP-pair production have a different asymptotic behaviour:
\begin{equation}
\frac{{\cal A}_{QCD}}{{\cal A}_{EW}}  \Big|_{s\to \infty} \sim   \frac{\ln s}{s}.
\label{eq:asympt}
\end{equation}

Therefore asymptotically the total cross section $\sigma_{\text{tot}} = \sigma_{EW}+\sigma_{int}+\sigma_{QCD}$ fall off with the increase of the energy as per
\begin{equation}\label{sig_full_by_s}
\sigma_{\text{tot}} \sim \frac{1}{s^2}\left(1 + {\cal O}\left(\frac{\ln s}{s}\right) 
\right),   
\end{equation}
where the main contribution proceeds from the tree level electroweak amplitude. It is interesting to note that the tree level QCD cross section of $J/\psi~\eta_c$-pair production falls off  with  energy increasing faster than (\ref{sig_full_by_s}), namely as $1/s^4$, because the latter process is  helicity suppressed (see \cite{Braaten:2002fi} for details).

\section{Cross sections estimations}\label{RES}

The numerical values of parameters used in the calculations are presented in  Table~\ref{tab:parameters}. The values of NRQCD matrix elements are adopted from the potential model~\cite{Eichten:2019hbb}.  The strong coupling constant is used within the two loops accuracy:
\begin{equation}\label{alphaS}
\alpha_S\left(Q\right) = \frac{4\pi}{\beta_0 L}\left(1-\frac{\beta_1\ln L}{\beta_0^2 L}\right),    
\end{equation} 
where $L= \ln{Q^{2}/\Lambda^{2}}$, $\beta_{0}= 11-\frac{2}{3}N_{f}$ and $\beta_{1}= 102-\frac{38}{3}N_{f}$ with $N_{f}= 5$; the reference value is $\alpha_S(M_{Z}) = 0.1179$. The appropriate scale $Q = \sqrt s$ is chosen for $\alpha_s$ for all investigated energies. For sake of simplicity  the fine structure constant is used in Thomson limit: $\alpha= 1/137$. 

As it is customary in most studies on quarkonia production within NRQCD, the masses of quarks inside the quarkonium are chosen so that their sum  is equal to the quarkonium mass.
 
\begin{table}[ht]
\centering
 \caption{The parameters used in the  calculations. The NRQCD matrix elements  are adopted from~\cite{Eichten:2019hbb}.}
\label{tab:parameters}
\begin{tabular}{cccc}
\hline
~~$m_c$ = 1.5~GeV~~  & ~~$m_b$ = 4.7~GeV~~  & ~~$M_Z$ = 91.2~GeV~~  & ~~$\Gamma_Z$ = 2.5~GeV~~ \\
\hline
\end{tabular}
\begin{tabular}{cccccc}
  $\langle O \rangle_{J/\psi} =\langle O \rangle_{\eta_c} = 0.523\mbox{ GeV}^3$ & & $\langle O \rangle_{\Upsilon} = \langle O \rangle_{\eta_b} = 2.797\mbox{ GeV}^3$ & & & $\sin^2\theta_w$ = 0.23  \\
\hline
\end{tabular}
\end{table}

\begin{figure}[ht]
    \centering
    \begin{minipage}[h]{0.49\linewidth} 
        \centering
        \includegraphics[width = 1\linewidth]{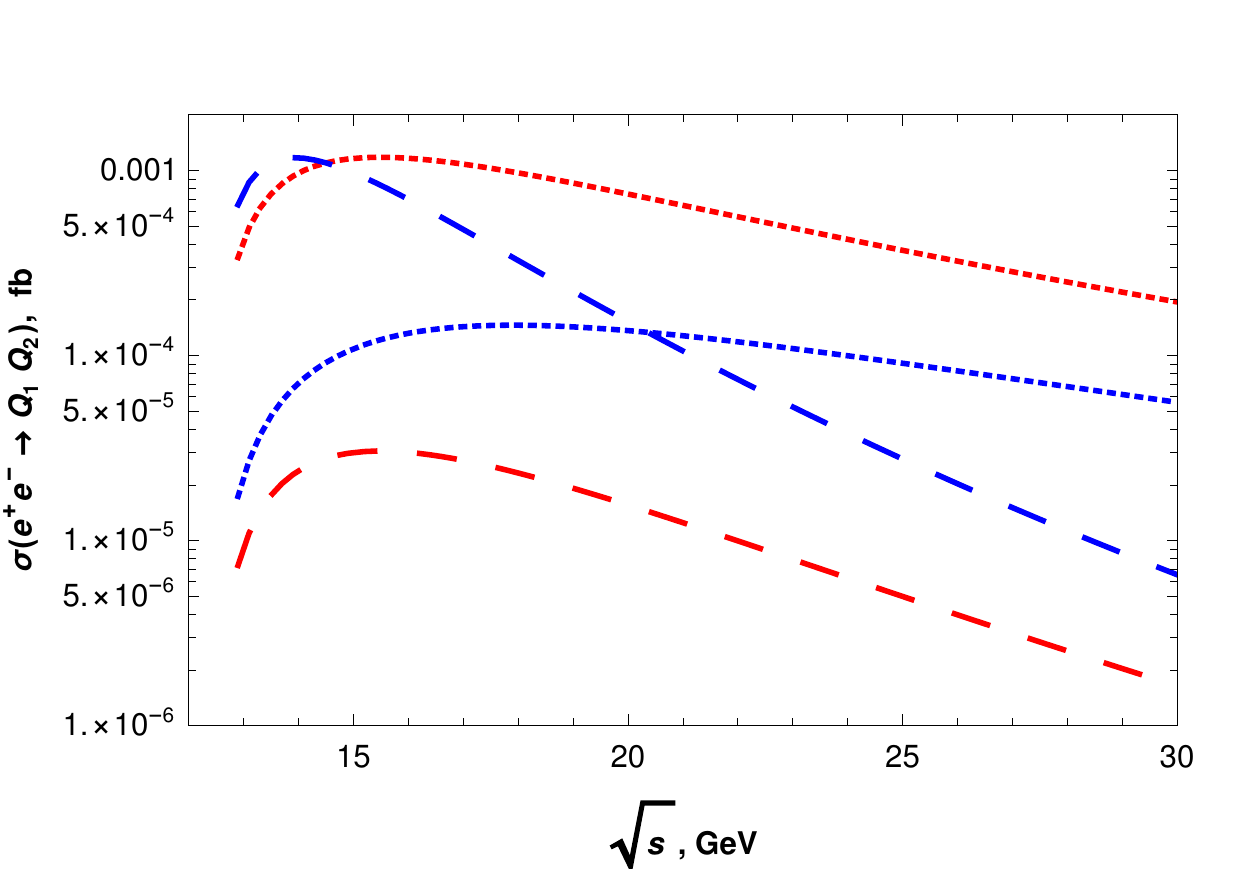}
    \end{minipage}
    \hfill
    \begin{minipage}[h]{0.49\linewidth} 
        \centering
        \includegraphics[width = 1\linewidth]{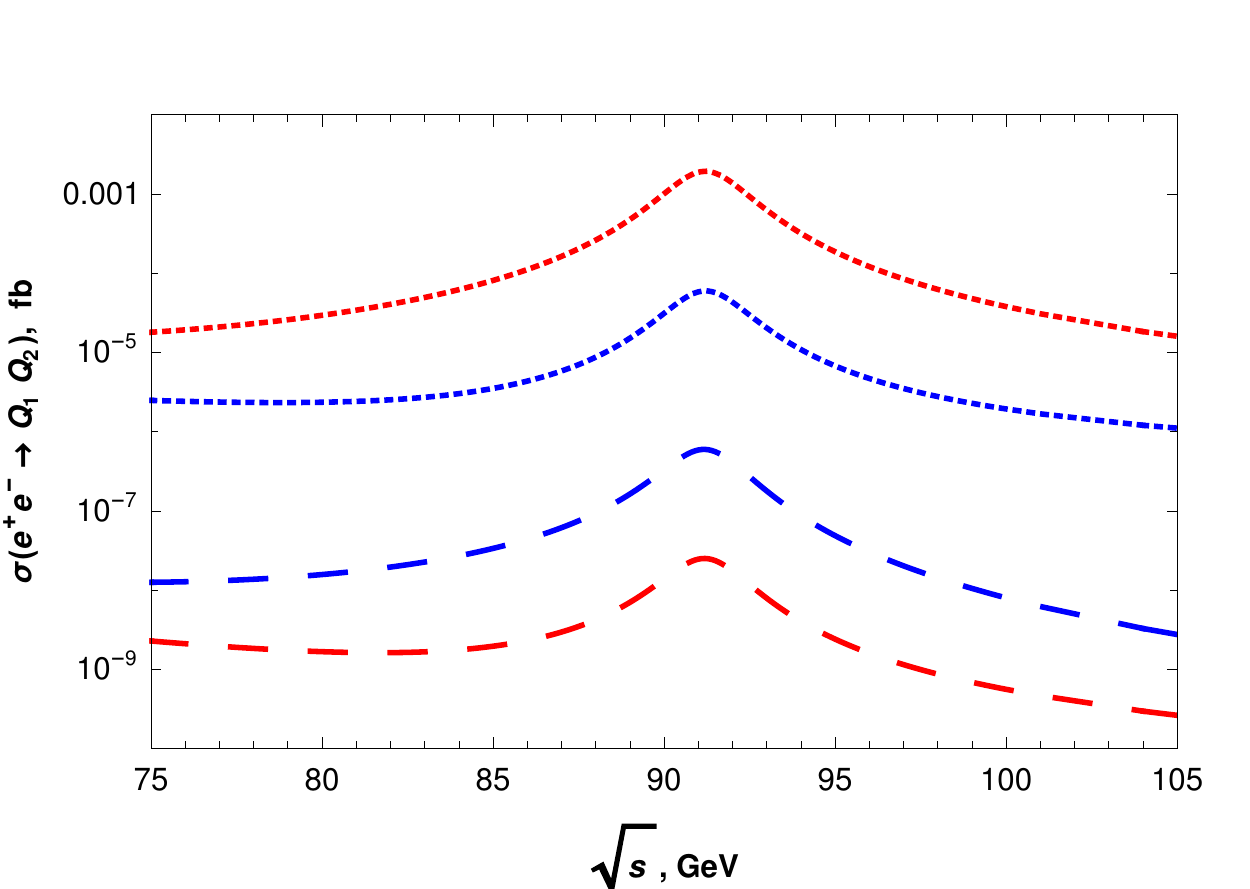}
    \end{minipage}
    \vfill
    \caption{The EW and QCD contributions to the cross sections at low energies (left) and near $Z$ pole (right): the QCD one loop contribution to $\sigma\left(J/\psi~\eta_b\right)$ (red dashed curve);  the EW contribution to $\sigma\left(J/\psi~\eta_b\right)$ (red dotted curve);   the QCD one loop contribution to $\sigma\left(\Upsilon~\eta_c\right)$ (blue dashed curve); the EW  contribution to $\sigma\left(\Upsilon~\eta_c\right)$ (blue dotted curve).}   
    \label{fig:sig_compare}
    \begin{minipage}[h]{0.9\linewidth}
        \includegraphics[width = \linewidth]{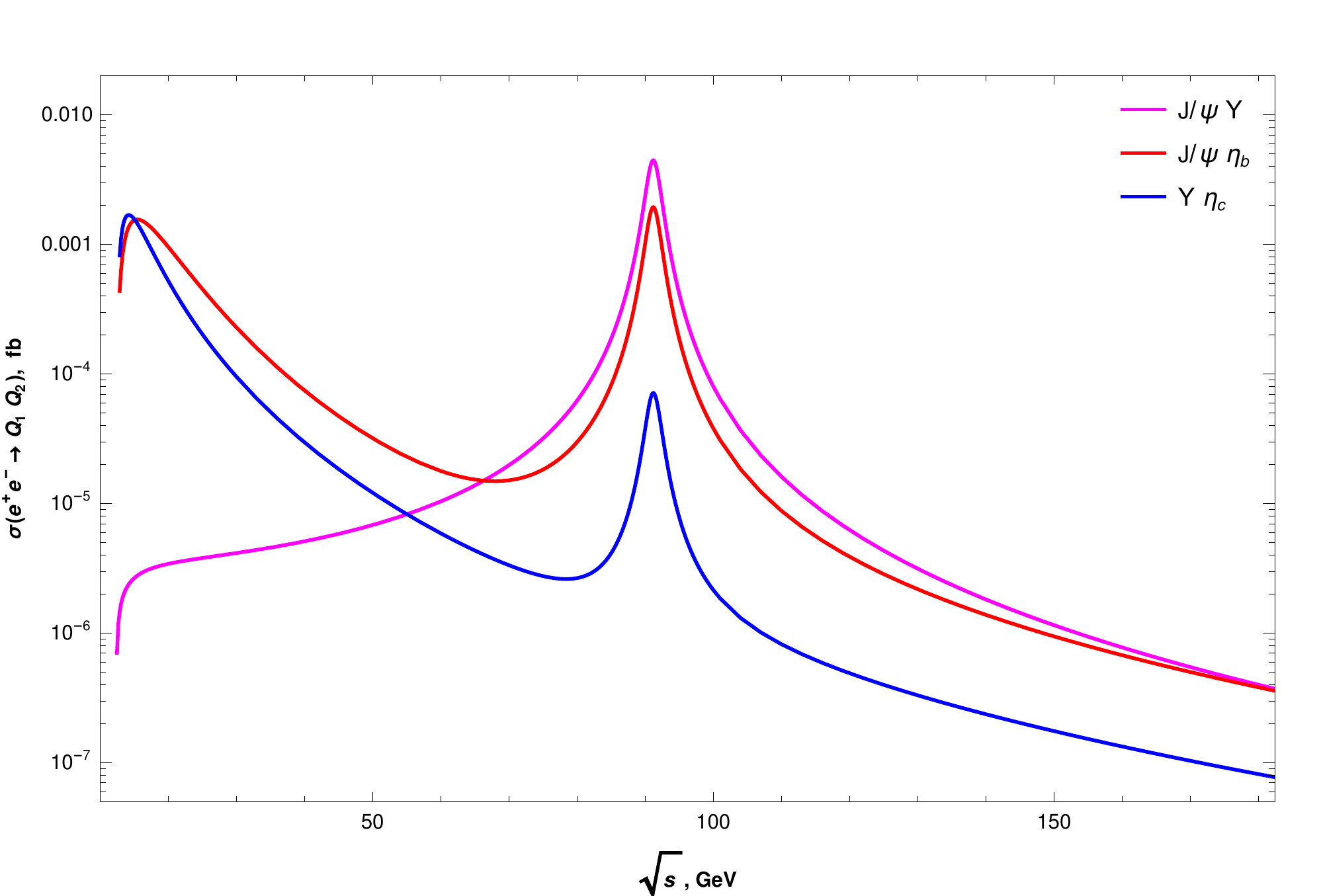}
        \caption{\centering The total cross sections dependence on the  collision energy.}
        \label{fig:sig_full}
    \end{minipage}
\end{figure}

\begin{figure}[ht]
    \begin{minipage}[h]{0.52\linewidth} 
        \centering
        \includegraphics[width = 1\linewidth]{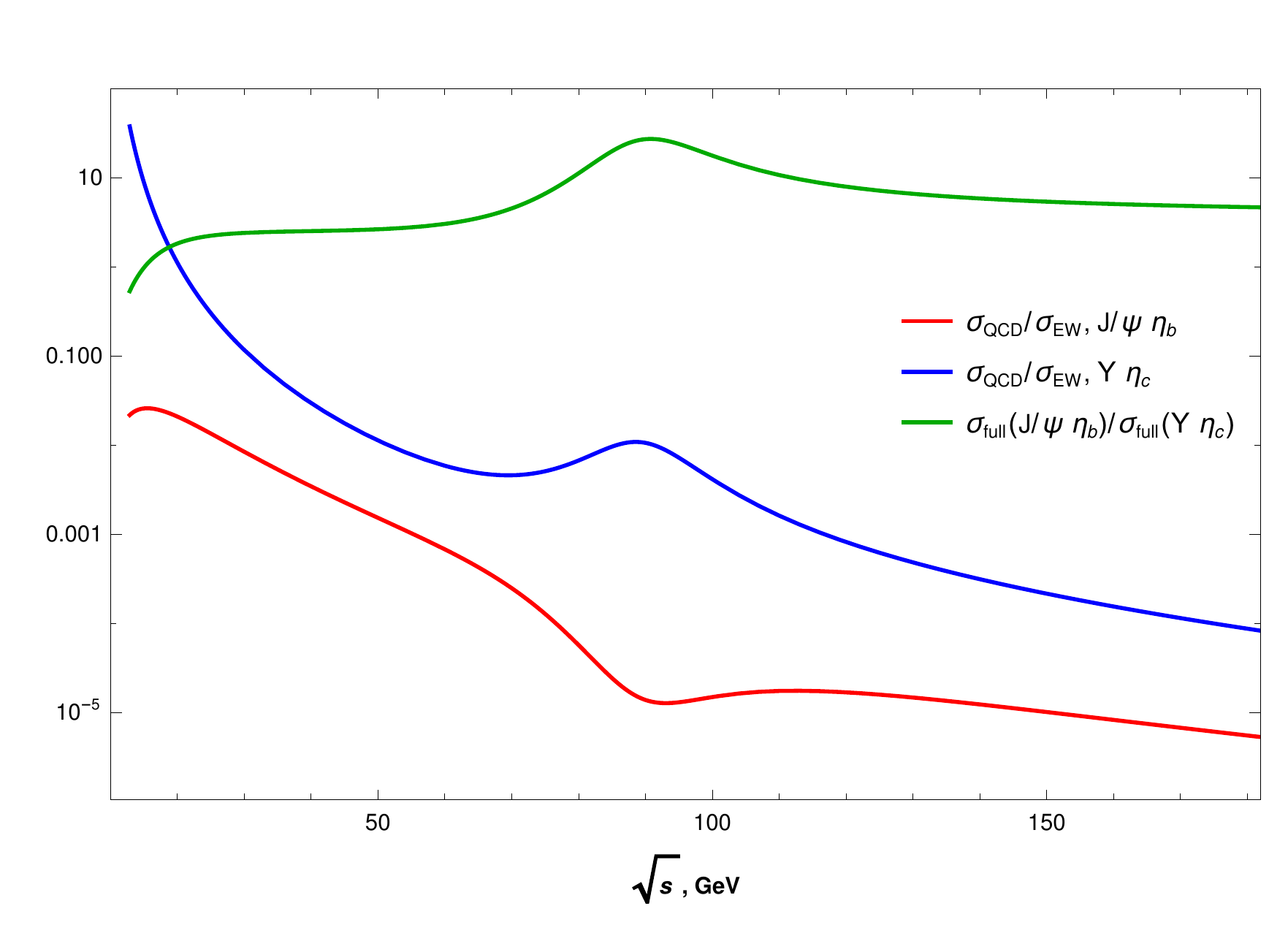}
\caption{
The cross section ratios as a function of the collision energy:  $\sigma_{QCD}(J/\psi\: \eta_b)/\sigma_{EW}(J/\psi\: \eta_b)$  (red curve), $\sigma_{QCD}(\Upsilon\: \eta_c)/\sigma_{EW}(\Upsilon\: \eta_c)$  (blue curve), $\sigma_{tot}(J/\psi\: \eta_b)/\sigma_{tot}(\Upsilon\: \eta_c)$ (green curve).  
}
        \label{fig:ratioQCD2EW}
        \vspace{22ex}
    \end{minipage}
    \hfill 
    \begin{minipage}[h]{0.4\linewidth}
        \centering
        \includegraphics[width = 1\linewidth]{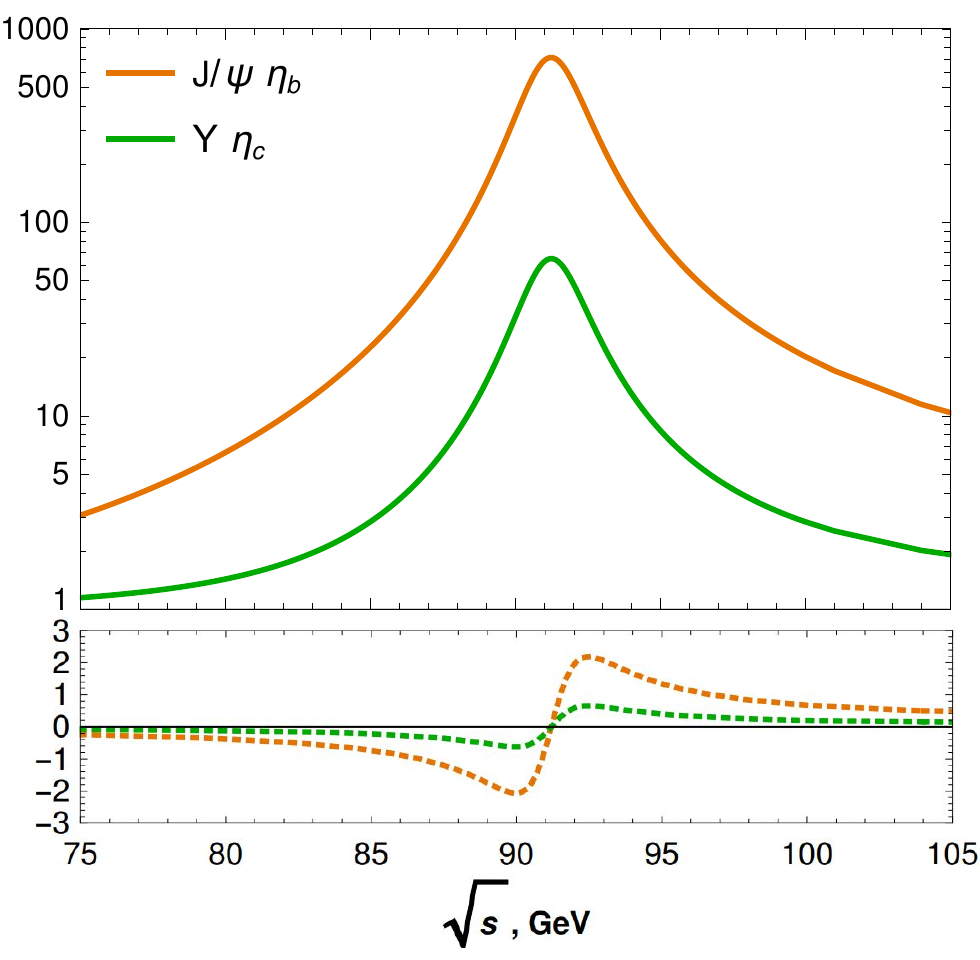}
        \caption{The $Z$-boson relative contributions as a function of energy: $\sigma_{\gamma+Z}(J/\psi\:
\eta_b)/\sigma_{\gamma}(J/\psi\: \eta_b)$ (solid orange curve) and $\sigma_{\gamma+Z}(\Upsilon\: \eta_c)/\sigma_{\gamma}(\Upsilon\: \eta_c)$ (solid green curve),  as well as the relative contributions of the interference term  between $\gamma$ and $Z$: $\sigma_{\gamma Z}^{int}(J/\psi\: \eta_b)/\sigma_{\gamma}(J/\psi\: \eta_b)$ (dashed orange curve) and $\sigma_{\gamma Z}^{int}(\Upsilon\: \eta_c)/\sigma_{\gamma}(\Upsilon\: \eta_c)$
       (dashed green curve).}
        \label{fig:ratioFull2gamma}
    \end{minipage}
\end{figure}

The cross sections reference values are given in Table~\ref{tab:cross-sec}. In Figures~\ref{fig:sig_compare},\ref{fig:sig_full},\ref{fig:ratioQCD2EW}, and \ref{fig:ratioFull2gamma} the calculated cross sections and there ratios are performed as functions of $\sqrt s$. 

In Figure~\ref{fig:sig_compare} we compare the  QCD and EW contributions at low energies (left) and at energies  around a $Z$-mass (right). 
In Figure~\ref{fig:sig_full} the total cross sections including all discussed contributions are demonstrated. In Figure~\ref{fig:ratioQCD2EW} the ratios between QCD and EW contributions are performed. 
The relative contribution of $Z$-boson annihilation to the studied processes is shown in Figure~\ref{fig:ratioFull2gamma}.

\begin{table}[ht]
\caption{The cross section values in fb units at different collision energies.} 
\centering
\label{tab:cross-sec}
\begin{tabular}{|c||c|c|c|c|c|c||}
\hline   
& $E=15~\text{GeV}$ & $E=20~\text{GeV}$ & $E=30~\text{GeV}$ & $E=50~\text{GeV}$ & $E=90~\text{GeV}$ & $E=180~\text{GeV}$  \\
\hline 
$~J/\psi~\eta_{b}~$  & $1.5\cdot10^{-3}$ & $9.5\cdot10^{-4}$ & $2.3\cdot10^{-4}$ & $3.2\cdot10^{-5}$ & $1.0\cdot10^{-3}$ & $3.8\cdot10^{-7}$ \\ 
$~\Upsilon~\eta_{c}~$ & $1.5\cdot10^{-3}$ &  $5.2\cdot10^{-4}$ & $9.6\cdot10^{-5}$ & $1.2\cdot10^{-5}$ & $3.7\cdot10^{-5}$ & $8.1\cdot10^{-8}$  \\
$~J/\psi~\Upsilon~$  & $2.6\cdot10^{-6}$ & $3.4\cdot10^{-6}$ & $4.1\cdot10^{-6}$  & $6.8\cdot10^{-6}$ & $2.3\cdot10^{-3}$ & $4.0\cdot10^{-7}$ \\
\hline\hline\hline
\end{tabular}
\end{table}

As it can be concluded from the presented Figures~\ref{fig:sig_compare},\ref{fig:sig_full},\ref{fig:ratioQCD2EW}, and \ref{fig:ratioFull2gamma}, the QCD and EW subprocesses contribute differently to the total yield of $J/\psi~\eta_b$ and $\Upsilon~\eta_c$. In the $J/\psi~\eta_b$ production  the  EW contribution dominates at all energies. Contrary, in the $\Upsilon~\eta_c$ production the EW contribution dominates only at high energies,  while at energies less than 20 GeV the main contribution comes from QCD mechanism.

To understand such a strange behaviour, we suggest to look at the problem from another side, and  compare the EW contributions with each other, as well as to compare the QCD contributions with each other. As it can be obtained  from the  expressions (\ref{eq:amp_ew_c}), (\ref{eq:amp_ew_b}), (\ref{eq:amp_qcd_c}) and (\ref{eq:amp_qcd_b}) for EW and QCD amplitudes, the scalar parts of the   amplitudes near the $Z$ pole relate as follows:
\begin{multline}
\frac{\mathcal{A}_{EW}^S\left(e^+e^-  \to J/\psi \: \eta_{b}\right)}{\mathcal{A}_{EW}^S\left(e^+e^-  \to \Upsilon \: \eta_{c}\right)}  \Big|_{s\sim m_Z\gg m_b,m_c} \approx \left[ \frac{\mathcal{A}_{QCD}^S\left(e^+e^-  \to J/\psi \: \eta_{b}\right)}{\mathcal{A}_{QCD}^S\left(e^+e^-  \to \Upsilon \: \eta_{c}\right)}  \Big|_{s\sim m_Z\gg m_b,m_c} \right]^{-1} \approx \\ \approx   \frac{(4e_b\sin^2\Theta_w+1)m_b}{(4e_c\sin^2\Theta_w-1)m_c}\approx -6.
\end{multline}

Thus, the ratios between the EW and QCD contributions  for the discussed processes  near the $Z$ pole are essentially different, and moreover,  these ratios are in inverse proportion to each other. Also it can be concluded from Figure~\ref{fig:sig_compare},  that the EW and QCD contributions approximately obey this pattern at all investigated energies:  $\sigma_{EW}\left(J/\psi\: \eta_b\right)$ is at least by an order of magnitude  greater, than $\sigma_{EW}\left(\Upsilon\: \eta_c\right)$, whereas $\sigma_{QCD}\left(J/\psi\: \eta_b\right)$ is at least by an order of magnitude smaller than $\sigma_{QCD}\left(\Upsilon\: \eta_c\right)$. If one keeps this circumstance in mind, then the behavior of the discussed contributions no longer seems so mysterious. Indeed, if at some energy 
$\sigma_{QCD}\left(\Upsilon~\eta_c\right) \sim \sigma_{EW}\left(\Upsilon~\eta_c\right)$, then at  this energy 
$\sigma_{EW}\left(J/\psi~\eta_b\right)\gg \sigma_{QCD}\left(J/\psi~\eta_b\right)$, because
$\sigma_{EW}\left(J/\psi~\eta_b\right)\gg \sigma_{EW}\left(\Upsilon~\eta_c\right)~$ and $~ \sigma_{QCD}\left(\Upsilon~\eta_c\right) \gg \sigma_{QCD}\left(J/\psi~\eta_b\right)$.

As  seen in Figure~\ref{fig:sig_full}, both the $J/\psi~\eta_b$-pair and the  $\Upsilon~\eta_c$-pair  production cross sections have a maximum near the threshold  ($\sqrt{s}_{max} (J/\psi~\eta_b) \approx 15.6~\text{GeV}$ and $\sqrt{s}_{max}(\Upsilon~\eta_c) \approx 14.1~\text{GeV}$).
The cross section ratios near the maximum take the following values:
\begin{equation}\label{ratios_low}
\frac{\sigma_{QCD}\left(J/\psi~\eta_b\right)}{\sigma_{EW}\left(J/\psi~\eta_b\right)} 
\sim 3 \cdot 10^{-2}, \qquad
\frac{\sigma_{QCD}\left(\Upsilon~\eta_c\right)}{\sigma_{EW}\left(\Upsilon~\eta_c\right)}
\sim 14, \qquad  
\frac{\sigma_{tot}\left(\Upsilon~\eta_c\right)}{\sigma_{tot}\left(J/\psi~\eta_b\right)} \sim 1.1 .
\end{equation}

As already mentioned, near the $Z$ pole the discussed cross sections behave completely differently: 
\begin{equation}\label{ratios_Z}
\frac{\sigma_{QCD}\left(J/\psi~\eta_b\right)}{\sigma_{EW}\left(J/\psi~\eta_b\right)} 
\sim  10^{-5}, \qquad
\frac{\sigma_{QCD}\left(\Upsilon~\eta_c\right)}{\sigma_{EW}\left(\Upsilon~\eta_c\right)}
\sim  10^{-2}, \qquad
\frac{\sigma_{tot}\left(\Upsilon~\eta_c\right)}{\sigma_{tot}\left(J/\psi~\eta_b\right)} \sim 4 \cdot 10^{-2}.
\end{equation}

As shown in Figure~\ref{fig:ratioQCD2EW}  
if the $\Upsilon~\eta_c$-pair is produced the EW contribution exceeds the QCD one starting with energy about $20~\text{GeV}$ which agrees with (\ref{eq:asympt}). It should be mentioned, that the  interference between  the EW and QCD contributions is strong and positive at all investigated energies. Particularly if the $\Upsilon~\eta_c$-pair is produced it achieves $\sim$48\% of the total cross section when the EW and QCD cross-sections are comparable.

Since $J/\psi~\Upsilon$ pair production goes only via the $Z$ boson exchange it is not surprising that such a process is highly  suppressed at low energies against the production of  $J/\psi~\eta_b$-pairs and   $\Upsilon~\eta_c$-pairs (see Figure~\ref{fig:sig_full}). 
However at energies higher than 70~GeV the production cross section of  $J/\psi~\Upsilon$ pair becomes greater than  the other cross sections: $\sigma\left(J/\psi~\Upsilon\right)>\sigma\left(J/\psi~\eta_b\right)>\sigma\left(\Upsilon~\eta_c\right)$. For example,  for the chosen parameter values  at $\sqrt{s}=M_Z$ we obtain that
\begin{equation}
\sigma\left(J/\psi~\Upsilon\right):\sigma\left(J/\psi~\eta_b\right):\sigma\left(\Upsilon~\eta_c\right) = 62.5 : 27.2 : 1. 
\end{equation}

The $Z$ boson exchange obviously dominates at the $Z$ pole,  and also essentially contributes to the production cross section around this pole is such a way that  the its contribution to the total cross section value is greater than 20\%  for energies
$\sqrt{s}>60~\text{GeV}$ for the  $J/\psi~\eta_b$-pair production process  and for energies in the range  $70~\text{GeV}< \sqrt{s}<150~\text{GeV}$ for 
 the  $\Upsilon~\eta_c$-pair production process (see Figure~\ref{fig:ratioFull2gamma}).




\section{Conclusions}
The  exclusive production of the charmonium-bottomonium pairs  ($J/\psi~\eta_b$,  $\Upsilon~\eta_c$ and $J/\psi~\Upsilon$) has been studied in a single boson $e^+e^-$~annihilation in the interaction energy range from the threshold to $2M_Z$  within the color singlet approximation of NRQCD.

Both  $J/\psi~\eta_b$  and $\Upsilon~\eta_c$ productions essentially differ from the thoroughly  investigated $J/\psi~\eta_c$ production, since  the main QCD contribution  to these processes contains loops and occurs to be  comparable in magnitude with the purely electromagnetic  contribution  (${\cal O}(\alpha^2 \alpha_s^4)$ v.s. ${\cal O}(\alpha^4)$). This is why the QCD contribution, the electromagnetic  contribution and their interference have been studied together. As concerned the $J/\psi~\Upsilon$-pair production,  in the  leading order this process goes via the electroweak $Z$ boson exchange only. The rather simple structure of the studied amplitudes allows one to provide the analytical expressions right in the text. 

It has been shown in the  current study, that the QCD and EW subprocesses contribute differently to the total yield of $J/\psi~\eta_b$ and $\Upsilon~\eta_c$. In the $J/\psi~\eta_b$ production  the  EW contribution obviously dominates at all energies. Contrary, in the $\Upsilon~\eta_c$ production the EW contribution dominates only at high energies, while at energies less than 20 GeV the main contribution comes from QCD mechanism. The suppression of one-loop QCD cross sections at high energies is explained by the fact that the one-loop QCD amplitude  and the electroweak amplitude possess a different asymptotic by $s$ powers.

The energy sufficient to produce a  charmonium-bottomonium pair can not be achieved at the current $e^+e^-$ experiments. Nevertheless we believe that the obtained  results may be of considerable interest for experiments at future $e^+e^-$ colliders.  

Authors would like to thank A.~Likhoded, A.~Onishchenko and S.~Poslavsky for help and fruitful discussions. The work was supported by foundation RFBR, grant No. 20-02-00154~A.~~I.~Belov acknowledges the support from ``BASIS'' Foundation, grant No. 20-2-2-2-1.

\clearpage
\appendix

\section{$\langle O \rangle$ and $|R(0)|^2$ values}
\label{APP1}
\setcounter{table}{0}
\renewcommand\thetable{\Alph{section}.\Roman{table}}
\begin{table}[h!]
\caption{$\langle O \rangle$ and $|R(0)|^2$ (where possible) in GeV$^3$ for $J/\psi$, $\eta_c$, $\Upsilon$ and $\eta_b$ mesons.}
\centering
\label{tab:nrqcd_me}
\begin{tabular}{|c|c|c||c|c|}
\hline
Ref. &
  $\langle O \rangle_{J/\psi}/ |R_{J/\psi}(0)|^2$ &
  $\langle O \rangle_{\eta_c}/ |R_{\eta_c}(0)|^2$ &
  $\langle O \rangle_{\Upsilon}/|R_{\Upsilon}(0)|^2$ &
  $\langle O \rangle_{\eta_b}/|R_{\eta_b}(0)|^2$ \\ \hline
\cite{Braaten:2002fi}  & $0.335 \pm 0.024$         & $0.297 \pm 0.032$          &                        &                       \\ \hline
\cite{Bodwin:2007fz}   & $0.440_{-0.055}^{+0.067}$ & $0.434 _{-0.158}^{+0.169}$ &                        &                       \\ \hline
\cite{Eichten:2019hbb} & \multicolumn{2}{c||}{$0.523/1.0952$}                    & \multicolumn{2}{c|}{$2.797/5.8588$}            \\ \hline
\cite{Chung:2010vz}    &                           &                            & \multicolumn{2}{c|}{$3.069^{+0.207}_{-0.190}$} \\ \hline
\end{tabular}
\end{table}

\section{Electroweak amplitudes with ${\cal O}\left(\frac{m_q^2}{M_Z^2-4m_q^2}\right)$ corrections }\label{APP2}

\begin{multline}\label{appendix_ew_c} 
\mathcal{A}_{EW}\left(e^+e^-\to J/\psi~\eta_b\right) = \frac{\langle O_{J/\psi} \rangle^{1/2}\langle O_{\eta_b} \rangle^{1/2} }{N_c}\times \\ \times
\frac{-3~e^4~e_be_c}{2 m_c\sqrt{m_b m_c}~\left(s+4m_c^2-4m_b^2\right)} \left(b_{\gamma}\left(1 + A_1\right)J_{\mu} + b_Z\left(1 + A_2\right)\widetilde{J}_{\mu}\right)\epsilon_{\nu}^{J/\psi}P_{\rho}Q_{\sigma}\varepsilon^{\mu\nu\rho\sigma}.
\end{multline}
\begin{multline}\label{appendix_ew_b} 
\mathcal{A}_{EW}\left(e^+e^-\to \Upsilon~\eta_c\right) = 
\frac{\langle O_{\Upsilon} \rangle^{1/2}\langle O_{\eta_c} \rangle^{1/2} }{N_c} \times \\ \times
\frac{3~e^4~e_be_c}{2 m_b\sqrt{m_b m_c}~\left(s+4m_b^2-4m_c^2\right)}\left(c_{\gamma}\left(1 + A_3\right)J_{\mu} + c_Z\left(1 + A_4\right)\widetilde{J}_{\mu}\right)\varepsilon_{\nu}^{\Upsilon}P_{\rho}Q_{\sigma}\varepsilon^{\mu\nu\rho\sigma}.
\end{multline}
\begin{multline}\label{appendix_ew_bc} 
{\cal A}_{EW}\left(e^+e^- \rightarrow J/\psi~\Upsilon\right) = \frac{\langle O_{J/\psi} \rangle^{1/2}\langle O_{\Upsilon} \rangle^{1/2} }{N_c}\ \times \\ \times \frac{3~e^4~e_be_c \sqrt{m_b m_c}}{\cos\theta_w\sin\theta_w \left(s-M_Z^2+i\Gamma M_Z\right)}
~\widetilde{J}_{\mu}\epsilon_{\nu_1}^{J/\psi}\epsilon_{\nu_2}^{\Upsilon}\varepsilon^{\mu \nu_1\nu_2\sigma}\times
\\
\times
\left(\frac{P_{\sigma}}{m_c^2 \left(4m_b^2-4 m_c^2-s\right)}\left(1 + A_5\right) - \frac{Q_{\sigma}}{m_b^2 \left(4 m_b^2-4
   m_c^2+s\right)}\left(1 + A_6\right)\right). 
\end{multline}

\small 
\begin{align*}
A_1 =  &-\left(\frac{m_c^2}{M_Z^2-4m_c^2}\right)\frac{\left(4e_b\sin^2\theta_w + 1 \right) \left(4e_c
\sin^2\theta_w - 1\right)}{4e_b e_c \cos^2\theta_w \sin^2\theta_w }, \\
A_2 = &-\left(\frac{m_c^2}{M_Z^2-4m_c^2}\right) \frac{\left(4 e_c \sin^2\theta_w-1\right)\left(\left(4 m_b^2 - 4m_c^2\right) \left(4 e_b \sin^2\theta_w + 1\right)^2 + s \left(8 e_b \sin^2\theta_w \left(2 e_b\sin^2\theta_w+1\right)+3\right)\right)}{4 e_b e_c \cos^2\theta_w \sin^2\theta_w\left(4 e_b \sin^2\theta_w+1\right)\left(4 m_b^2-4 m_c^2+s\right)},\\ 
A_3 = &-\left(\frac{m_b^2}{M_Z^2-4m_b^2}\right) \frac{\left(4 e_b \sin^2\theta_w + 1\right) \left(4 e_c \sin^2\theta_w-1\right)}{4 e_b e_c\cos^2\theta_w \sin^2\theta_w},\\
A_4 = &-\left(\frac{m_b^2}{M_Z^2-4m_b^2}\right)\frac{\left(4 e_b \sin^2\theta_w +1\right)\left(\left(4m_b^2 - 4 m_c^2\right)\left(4 e_c\sin^2\theta_w - 1\right)^2 - s\left(8e_c\sin^2\theta_w (2e_c\sin^2\theta_w - 1) + 3\right) \right)}{4 e_b e_c \cos^2\theta_w \sin^2\theta_w\left(4 e_c \sin^2\theta_w-1\right)\left(4 m_b^2-4 m_c^2-s\right)}, \\
A_5 = &-\left(\frac{m_c^2}{M_Z^2-4m_c^2}\right)\frac{\left(4 e_b \sin^2\theta_w+1\right) \left(4 e_c \sin^2\theta_w-1\right) \left(4m_b^2 - 4m_c^2+3 s\right)}{4 e_be_c \cos^2\theta_w \sin^2\theta_w \left(4 m_b^2-4 m_c^2+s\right)}, \\    
A_6 = &-\left(\frac{m_b^2}{M_Z^2-4m_b^2}\right)\frac{\left(4 e_b \sin^2\theta_w+1\right) \left(4 e_c \sin^2\theta_w-1\right) \left(4 m_b^2-4 m_c^2-3 s\right)}{4 e_b e_c \cos^2\theta_w \sin^2\theta_w \left(4 m_b^2-4 m_c^2-s\right)}.    
\end{align*}
\normalsize



\vspace{3ex}
\section*{References}

\end{document}